\documentclass[
 amsmath,amssymb,
preprint]{revtex4-1}

\usepackage{graphicx}
\usepackage{dcolumn}
\usepackage{bm}
\usepackage[normalem]{ulem} 
\usepackage[utf8]{inputenc}
\usepackage[T1]{fontenc}
\usepackage{mathptmx}
\usepackage{etoolbox}
\usepackage{color}
\newcommand{\tr}[1]{\textcolor{red}{#1}}

\makeatletter
\def\@email#1#2{%
 \endgroup
 \patchcmd{\titleblock@produce}
  {\frontmatter@RRAPformat}
  {\frontmatter@RRAPformat{\produce@RRAP{*#1\href{mailto:#2}{#2}}}\frontmatter@RRAPformat}
  {}{}
}%
\makeatother

\begin{document}

\title[]{
Voltage controlled iontronic switches: a computational method to predict electrowetting in hydrophobically gated nanopores}
\author{Gon\c{c}alo Paulo}
\author{Alberto Gubbiotti}%
\author{Giovanni Di Muccio}
\author{Alberto Giacomello$^\ast$}
\email{alberto.giacomello@uniroma1.it}
\affiliation{Dipartimento di Ingegneria Meccanica e Aerospaziale, Sapienza Universit\`a di Roma, Rome, Italy}

\date{\today}

\begin{abstract}
Reliable and controllable switches are crucial in nanofluidics and iontronics. Ion channels in nature serve as a rich source of inspiration due to their intricate mechanisms modulated by stimuli like pressure, temperature, chemicals, and voltage. The artificial replication of the properties of these channels is challenging due to their complex chemistry, limited stability range, and intricate moving parts, allosterically modulated. Nonetheless, we can harness some of their gating mechanisms for nanofluidic and iontronic purposes. This theoretical and computational study explores the use of electrowetting in hydrophobic nanopores to control their conductance using an external applied voltage. We employ restrained molecular dynamics to calculate the free energy required for wetting a model nanopore under different voltages. Utilizing a simple theory, we generate free energy profiles across a wide voltage range. We also computed transition rates between conductive and non-conductive states, showing their voltage dependence and how this behaviour can impair memory to the system, resembling the memristor behaviour voltage-gated channels in the brain. These findings offer a promising avenue for designing and controlling hydrophobic nanopores via electrowetting, enabling potential applications in neuromorphic iontronics.
\end{abstract}

\maketitle

\section{Introduction}

Nature has long been a wellspring of inspiration for scientific innovation, offering intricate designs and dynamic processes that have fueled advances across various domains~\cite{vincent2006biomimetics,bhushan2009biomimetics}.
It can demonstrate some of the best examples of both iontronics~\cite{arbring2018decade} -- exemplified by the propagation of the action potential~\cite{gidon2020dendritic} -- and nanofluidics, by presenting us with the complex machinery that are ion channels and many other biological pores,  capable of a highly specialized modulation of the signalling and transport of ions, water and biological molecules across the cell membranes~\cite{hille1970ionic,macaulay2021molecular}.

The complex transport proprieties of such membrane proteins are modulated by their shape and geometrical features~\cite{lynch2020water,roux2004,corry2006};
e.g., the cell membrane potential is used in many cells to regulate their physiological functions, in particular regulating the permeability to sodium and potassium ions in nerve and muscle tissues~\cite{almers2005gating}. The voltage dependence of Na+ and K+ conductances requires that the channels have a sensor, 
such as a charge or a dipole that responds to voltage changes~\cite{bezanilla2008ion, hodgkin1952quantitative}.
Indeed, voltage-gated ion channels use the movement of complex voltage sensing domains to induce configuration changes, usually generating steric occlusion of the channel~\cite{bassetto2023noncanonical, furini2012conduction}.
These channels have been a primary source of inspiration for neuromorphic computing~\cite{Xiong2023,Najem2018,Indiveri2011,Zhu2020}, 
as they are effectively memristors~\cite{Xu2019,Hegab2015,CHUA2012}. 
However, despite the great advancement of protein engineering~\cite{brannigan2002protein,huang2016coming, zhao2019electro}, 
the precision engineering of proteins with moving part (and related allosteric regulation) remains a formidable challenge.

Here we focus on hydrophobically gated nanopores, that can modulate their transport thanks to the formation of vapour bubbles inside their lumen~\cite{Aryal2015}.
This physical gating mechanism is a rather simple and reliable way to control the current through the pore that can be modulated by an external electric field.
Indeed, despite wetting of hydrophobic cavities is usually induced by the application of an external pressure~\cite{giacomello2023}
it is well established that an electrical field can decrease the apparent contact angle of water on top of a hydrophobic material~\cite{lippmann1875relations,mugele2005}
and that voltage can favor wetting of hydrophobic pores~\cite{Dzubiella2005,vanzo2014,klesse2020,powell2011,smirnov2011,Polster2022}.
Recently, we have also shown that hydrophobic gating and electrowetting can be combined to develop memristive applications~\cite{paulomem2023}.
Nevertheless, no conclusive explanation or theory has been established on the effect of voltage on the wetting properties of narrow pores.
By using molecular dynamics simulations and enhanced sampling techniques, we developed a theoretical framework to predict the free energy profile related to the filling of an hydrophobic pore under an applied voltage, as well as a theoretical approach that in principle could be used to predict the wetting free energy profile at any voltage, starting from the one computed at zero voltage. This quantitative tools can be used to understand the coupling of voltage and wetting in hydrophobically gated nanopores and to design optimized nanofluidic memristors.

\section{Methodology}
\subsection{Molecular dynamics setup}

\begin{figure}
    \centering
    \includegraphics[width=1\textwidth]{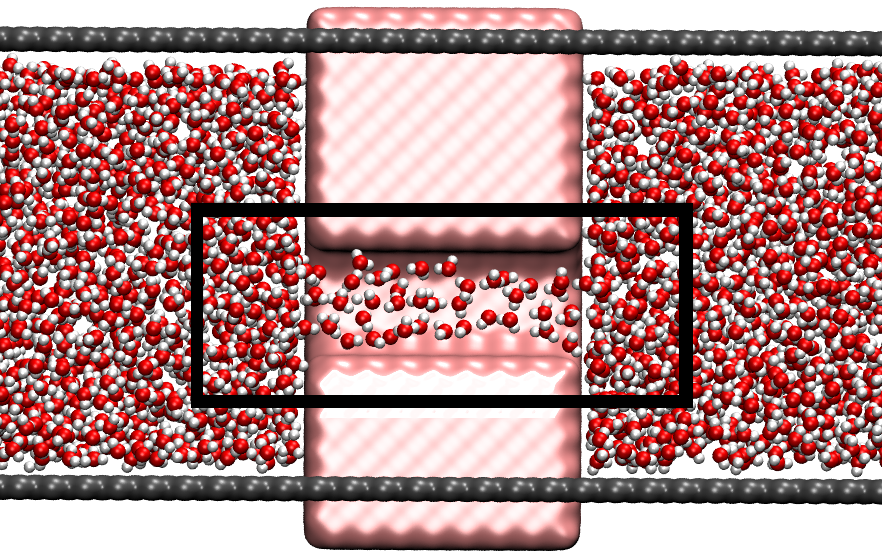}
	\caption{System setup. 
 Atomistic system used in this study, consisting of two reservoirs of liquid SPC/E water separated by a cylindrical nanopore drilled into a membrane formed by Lennard-Jones atoms.
 The lateral pistons, represented in black, composed by Lennard-Jones atoms also, are used to control the pressure by applying a force on water molecules corresponding to 1~atm. The black rectangle enclosing the pore represents the control region used to compute the water filling $N$ and to apply the filling constrain. 
 The picture is realized with the VMD software~\cite{vmd}.}
    \label{fig:system}
\end{figure}

Following the same protocol of our previous work~\cite{Paulo2023JCP}, we built a molecular dynamics system for which we want to compute the free energy $F$ and the diffusivity $D$ associated with the number of water molecules $N$ inside a hydrophobic nanopore; $N$ represents the coarse-grained variable which defines the wetting/drying process.
The system, represented in Fig.~\ref{fig:system}, is made of a slab of fixed Lennard-Jones atoms in a fcc arrangement, with lattice spacing 0.35 $nm$, from which a cylindrical nanopore was excavated. This slab is surrounded by water molecules (SPC/E~\cite{berendsen1987missing}), and the water-solid non-bonded interactions are  tuned~\cite{camisasca2020gas} so that the contact angle is $104^\circ$.
The nanopore has a diameter of 1.4~$nm$ and a length of 2.8~$nm$.
To control the pressure of the system, we used two pistons orthogonal to the pore axis~\cite{marchio2018}. The NVT ensemble was sampled using a Nos\'e--Hoover chains thermostat~\cite{martyna1992} at 310~K with a chain length of 3. A constant and homogeneous electric field is applied across the system $\mathrm{E}=(0,0,E_z)$, with $z$ being the direction parallel to the pore axis, to mimic a difference of a constant voltage across the membrane $\Phi=E_z L_z$~\cite{gumbart2012constant}, $L_z$ the length of the MD box.

\subsection{Free energy and diffusivity computation}
We use Restrained Molecular Dynamics (RMD)~\cite{maragliano2006temperature} to compute the free energy as a function of the pore filling $N$.
This is done by adding a harmonic restraint to the physical Hamiltonian $H_\mathrm{phys}$  of the system,
\begin{equation}\label{eq:restraint}
    H_N(\boldsymbol{r},\boldsymbol{p})=H_\mathrm{phys}(\boldsymbol{r},\boldsymbol{p})+\frac{k}{2}\left(N-\tilde{N}(\boldsymbol{r})\right)^2\,,
\end{equation}
where $\boldsymbol{r}$ and $\boldsymbol{p}$ are the positions and momenta of all the atoms, respectively, $k$ is a harmonic constant (1~kcal/mol), $N$ is the desired number of water molecules in a box centered around the nanopore, and $\tilde{N}$ is computed by counting the number of water molecules in the region highlighted in black in Fig.~\ref{fig:system}. 
The counting procedure has been explained in detail in previous work~\cite{Paulo2023JCP}.
The diffusivity $D(N)$, associated to the filling variable, can be computed within the same framework, being related to the integral of the autocorrelation function of $\tilde{N}$ at a given $N$~\cite{Paulo2023JCP}.

\section{Theory}
\subsection{Dependence of the free energy on the applied voltage}

In this section, we explore the effect that a transmembrane voltage has on the free energy profile associated with a given collective variable $\theta$, which in the specific case considered here is the number of water molecules inside the pore, $\theta = N$.
Here, the main assumption is that the system is at equilibrium even when applying the voltage.
Consider a system described by the vector $\boldsymbol{x}$ of positions and momenta of its $M$ particles
$\boldsymbol{x}=(\boldsymbol{r_1},\boldsymbol{p_1},\dots,\boldsymbol{r_M},\boldsymbol{p_M})$,
with a Hamiltonian made by a ‘‘zero voltage'' part $H_0$ and the contribution of the external voltage $\Phi$
\begin{equation}
H(\boldsymbol{x},\Phi)=H_0(\boldsymbol{x})-Q(\boldsymbol{r})\Phi\;,
\end{equation}
with $Q$ the displacement charge
\begin{equation}
Q(\boldsymbol{r})=\frac{\pi(\boldsymbol{r})}{L}\;,
\end{equation}
where
\begin{equation}
\pi(\boldsymbol{r})=\sum\limits_{i=1}^M q_i\boldsymbol{r}_i\cdot\boldsymbol{\hat{n}}
\end{equation}
is the total dipole moment of the system in the direction $\boldsymbol{\hat{n}}$ in which the voltage is applied
and $L$ is the extension of the system in the same direction.

The probability density function of the system in the canonical ensemble is then
\begin{equation}
\rho(\boldsymbol{x},\Phi)=\frac{1}{Z(\Phi)}e^{-\beta H(\boldsymbol{x},\Phi)}\;,
\end{equation}
with $Z$ the partition function
\begin{equation}
Z(\Phi)=\int e^{-\beta H(\boldsymbol{x},\Phi)}\mathrm{d}\boldsymbol{x}\;.
\end{equation}

Choosing a collective variable $\theta$ which gives a coarser description of the system, 
the probability density of that variable is 
\begin{equation}
P(\theta,\Phi)=\int \delta\left(\theta - \hat{\theta}(\boldsymbol{x})\right)\rho(\boldsymbol{x},\Phi)\mathrm{d}\boldsymbol{x}=\frac{Z_\theta(\theta,\Phi)}{Z(\Phi)}\;,
\end{equation}
where $\hat{\theta}(\boldsymbol{x})$ is the function which associates the microscopic state to a value of the collective variable, and $Z_\theta$ is the partition function restrained to a given value of the collective variable 
\begin{equation}
Z_\theta(\theta,\Phi)=\int \delta\left(\theta-\hat{\theta}(\boldsymbol{x})\right)e^{-\beta H(\boldsymbol{x},\Phi)}\mathrm{d}\boldsymbol{x}\;.
\end{equation}
The associated free energy is
\begin{equation}
F(\theta,\Phi)=-\beta^{-1}\log P(\theta,\Phi)\;,
\end{equation}
and its variation with respect to voltage 
\begin{equation}
\frac{\partial F(\theta,\Phi)}{\partial \Phi}=-\frac{1}{\beta P(\theta,\Phi)}\frac{\partial P(\theta,\Phi)}{\partial \Phi}\;.
\end{equation}
The derivative of the coarse grained probability with respect to voltage is
\begin{equation}
\frac{\partial P(\theta,\Phi)}{\partial \Phi}=\int \delta\left(\theta-\hat{\theta}(\boldsymbol{x})\right)\frac{\partial \rho(\boldsymbol{x},\Phi)}{\partial \Phi}\mathrm{d}\boldsymbol{x}\;,
\end{equation}
with
\begin{equation}
\frac{\partial \rho(\boldsymbol{x},\Phi)}{\partial \Phi}=\left(\frac{\beta Q(\boldsymbol{r})}{Z(\Phi)}-\frac{\mathrm{d}Z}{\mathrm{d}\Phi}\frac{1}{Z(\Phi)^2}\right)e^{-\beta H(\boldsymbol{x},\Phi)}
\end{equation}
and 
\begin{equation}
\frac{\mathrm{d} Z(\Phi)}{\mathrm{d}\Phi}=\int \beta Q(\boldsymbol{r})e^{-\beta H(\boldsymbol{x},\Phi)}\mathrm{d}\boldsymbol{x}=\beta Z(\Phi)\left\langle Q\right\rangle_\Phi\;.
\end{equation}
Hence, we obtain an expression for the derivative of the microscopic probability density with respect to voltage
\begin{equation}\label{eq:pdf_derivative}
\frac{\partial \rho(\boldsymbol{x},\Phi)}{\partial \Phi}=\frac{\beta}{Z(\Phi)}\left( Q(\boldsymbol{r})-\left\langle Q\right\rangle_\Phi\right)e^{-\beta H(\boldsymbol{x},\Phi)}\;,
\end{equation}
which can be used to compute the derivative for the coarse grained probability density
\begin{equation}
\frac{\partial P(\theta,\Phi)}{\partial \Phi}=\int \delta\left(\theta-\hat{\theta}(\boldsymbol{x})\right)\frac{\beta}{Z(\Phi)}\left( Q(\boldsymbol{r})-\left\langle Q\right\rangle_\Phi\right)e^{-\beta H(\boldsymbol{x},\Phi)}\mathrm{d}\boldsymbol{x}\;,
\end{equation}
simplifying
\begin{equation}
\frac{\partial P(\theta,\Phi)}{\partial \Phi}=\beta P(\theta, \Phi)\left(\left\langle Q\right\rangle_{\theta,\Phi}-\left\langle Q\right\rangle_\Phi\right)\;.
\end{equation}

We finally obtain an expression for the first order derivative of the free energy with respect to voltage
\begin{equation}
\frac{\partial F(\theta,\Phi)}{\partial \Phi}=\left\langle Q\right\rangle_\Phi-\left\langle Q\right\rangle_{\theta,\Phi}
\end{equation}

To obtain the second order derivative,
\begin{equation}
\frac{\partial F(\theta,\Phi)^2}{\partial^2 \Phi}=\frac{\partial\left\langle Q\right\rangle_\Phi}{\partial \Phi}-\frac{\partial\left\langle Q\right\rangle_{\theta,\Phi}}{\partial \Phi}\;,
\end{equation}
we need to compute the derivative of the expected averages. Consider the generic observable $A$, then:
\begin{equation}
\frac{\partial \left\langle A\right\rangle_\Phi}{\partial \Phi}=\int A(\boldsymbol{x})\frac{\partial\rho(\boldsymbol{x},\Phi)}{\partial \Phi}\mathrm{d}\boldsymbol{x}\;,
\end{equation}
we can use Eq.~\eqref{eq:pdf_derivative}
\begin{equation}
\frac{\partial \left\langle A\right\rangle_\Phi}{\partial \Phi}=\int \frac{\beta A(\boldsymbol{x})}{Z(\Phi)}\left( Q(\boldsymbol{r})-\left\langle Q\right\rangle_\Phi\right)e^{-\beta H(\boldsymbol{x},\Phi)}\mathrm{d}\boldsymbol{x}\;,
\end{equation}
to obtain eventually
\begin{equation}
\frac{\partial \left\langle A\right\rangle_\Phi}{\partial \Phi}=\beta\left(\left\langle AQ\right\rangle_\Phi-\left\langle A\right\rangle_\Phi\left\langle Q\right\rangle_\Phi\right)\;.
\end{equation}

Now we can compute the second order derivative of the free energy with respect to voltage
\begin{equation}
\frac{\partial F(\theta,\Phi)^2}{\partial^2 \Phi}=\beta\left(\left\langle Q^2\right\rangle_\Phi-\left\langle Q\right\rangle^2_\Phi - (\left\langle Q\right\rangle^2_{\theta,\Phi}-\left\langle Q^2\right\rangle_{\theta,\Phi}) \right)\;.
\end{equation}

The free energy dependence on the voltage can be then approximated, up to second order, as:

\begin{equation}
    F(\theta,\Phi) = F_0(\theta,\Phi_0) +
                     P_1(\theta,\Phi_0) \Delta \Phi +
                     \frac{1}{2}P_2(\theta,\Phi_0) (\Delta \Phi)^2 +
                     const(\Phi)
                     \;,
    \label{eq:theory_free}
\end{equation}
where $\Delta \Phi = (\Phi-\Phi_0)$, 
with $\Phi_0$ being voltage around which the expansion is done;
$P_1(\theta,\Phi_0) =-\left\langle Q\right\rangle_{\theta,\Phi_0}$ and
$P_2(\theta) = -\beta\left(\left\langle Q^2\right\rangle_{\theta,\Phi_0}-\left\langle Q\right\rangle^2_{\theta,\Phi_0}\right)$;
in the constant we collected all the terms that do not depend on $\theta$,
$const(\Phi) = 
\left\langle Q \right\rangle_{\Phi_0} \Delta \Phi + 
\left( \left\langle Q^2 \right\rangle_{\Phi_0} - 
\left\langle Q \right\rangle^2_{\Phi_0} \right) 
\left(\Delta \Phi \right)^2$.
While this framework is valid for a generic collective variable $\theta$, in the following, we will focus only on the case where the collective variable is  the pore filling $N$. During the restrained molecular dynamics simulations used to compute the free energy profiles, we computed also the values of $P_1$ and $P_2$ for different filling levels and different applied voltages.

\section{Results}
\subsection{Average displacement charge and second moment of the displacement charge at 0 V}

\begin{figure*}[t]
    \centering
    \includegraphics[width=\linewidth]{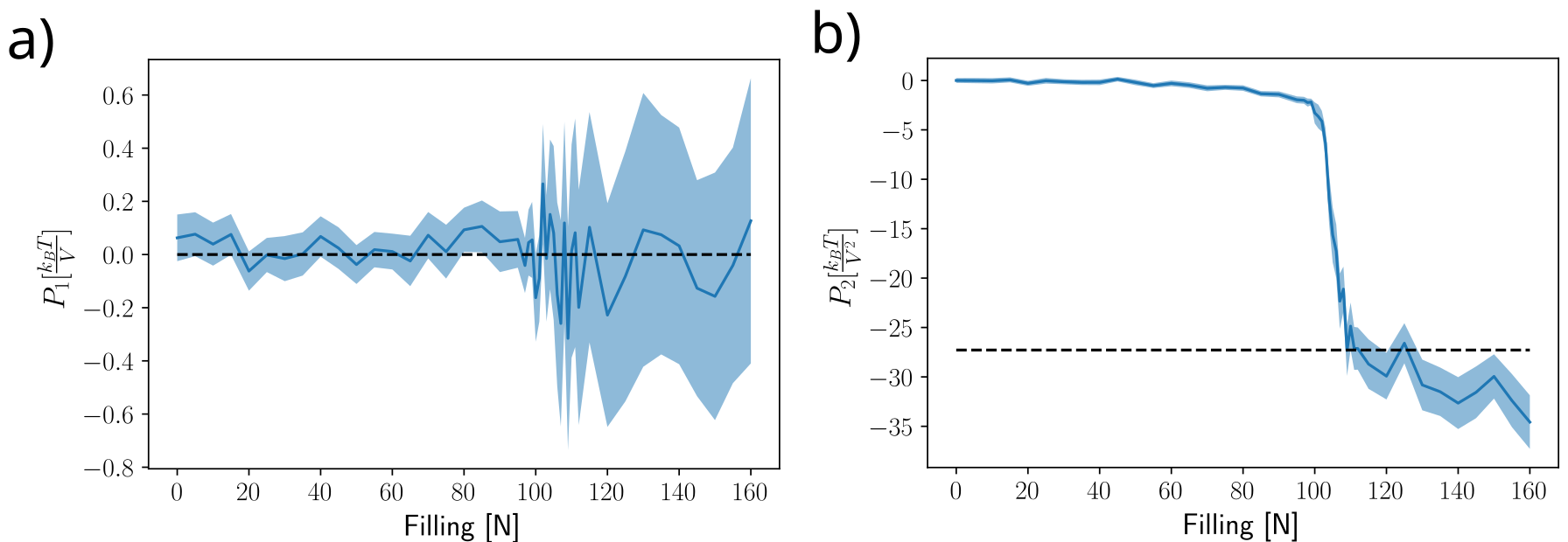}
    \caption{\textbf{Values of the first order coefficient, $P_1$, and the second order coefficient, $P_2$.} 
    During the restrained molecular dynamics simulations, we computed the first and second order coefficients $P_1$ and $P_2$ of Eq.~\eqref{eq:theory_free} at different pore fillings, $N$. 
    Because of the symmetry of the system, $P_1$ is almost constant, since opposite voltages must have the same effect. $P_2$ has the shape of a sigmoid function, panel b, since the fluctuations of total dipole $Q$ can be somehow related to the electric permittivity of the dielectric medium filling the pore, see the text for a detailed explanation. $P_1$ and $P_2$ are shifted to zero at $N=0$ as the free energy is always defined up to a constant.
    Shaded areas correspond to the 95\% confidence interval computed by boostraping the averages and variances of different blocks of measurements of Q.
    \label{fig:displacementcharge}
    }
\end{figure*}

During the restrained molecular dynamics simulations used to compute the free energy, we computed the average displacement charge $\langle Q \rangle_{\theta,0}$  and its second moment,$-\left(\left\langle Q^2\right\rangle_{\theta,\Phi_0}-\left\langle Q\right\rangle^2_{\theta,\Phi_0}\right)$, which we then use to compute $P_1$ and $P_2$ in Eq.~\eqref{eq:theory_free} as a function of the water filling of the pore; data are reported in Fig.~\ref{fig:displacementcharge}.
Because the pore shown in Fig.~\ref{fig:system} is i) symmetric, ii) uncharged, and iii) the displacement charge enters as a factor proportional to the voltage, $P_1$ should not depend on the filling of the pore, as positive and negative voltages must have the same effect on wetting and drying.
This argument is backed up by the MD results showing that indeed $P_1$ fluctuates around a constant value (Fig.~\ref{fig:displacementcharge}a). 
Differently, the amplitude of the fluctuation of $P_1$ sharply increases after a certain filling level, see panel b reporting the value of the second order coefficient $P_2$.
We related this effect to the fact that the number of fluctuating dipoles (the water molecules) inside the pore is increasing, passing from nearly vacuum (dry) to a much complex situation in which a larger number of molecules is present that can be potentially affected by the external electric field (the applied voltage drop would mainly occur inside the nanopore).
Indeed, $P_2$ has the dimensions of a capacitance, so the second term added to the free energy could be related to the energy stored in the pore, considering it as a capacitor. In particular, the change in the stored energy depends on the filling medium: the wet pore is filled with water (relative permittivity $\epsilon_r \gg 1 $), while the dry pore is filled with vapour (i.e., for such confined geometry, vacuum with $\epsilon_r \sim 1$). Hence, the capacitance of the dry pore has to be lower than the wet pore; we show by a dashed line what should be the energy of a capacitor with the size of the pore and a relative permittivity $\epsilon_r$ of c.a. 40; the latter value is lower than the bulk value due to the extreme confinement, in agreement with what has been reported for pores of  sizes similar to the one studied here~\cite{Dzubiella2005}.

The fact that $P_1$ is constant and only $P_2$ depends on the filling level implies that the wetting probability of {\it symmetric} hydrophobic nanopores would depend only on the square of the applied voltage, and not on any linear term, because it would be the only influence on the wetting free energy, in line with what has been reported by other authors~\cite{Trick2017}.

\subsection{Effect of applied voltage on the free energy and diffusivity as a function of pore filling}

\begin{figure*}
    \centering
    \includegraphics[width=\linewidth]{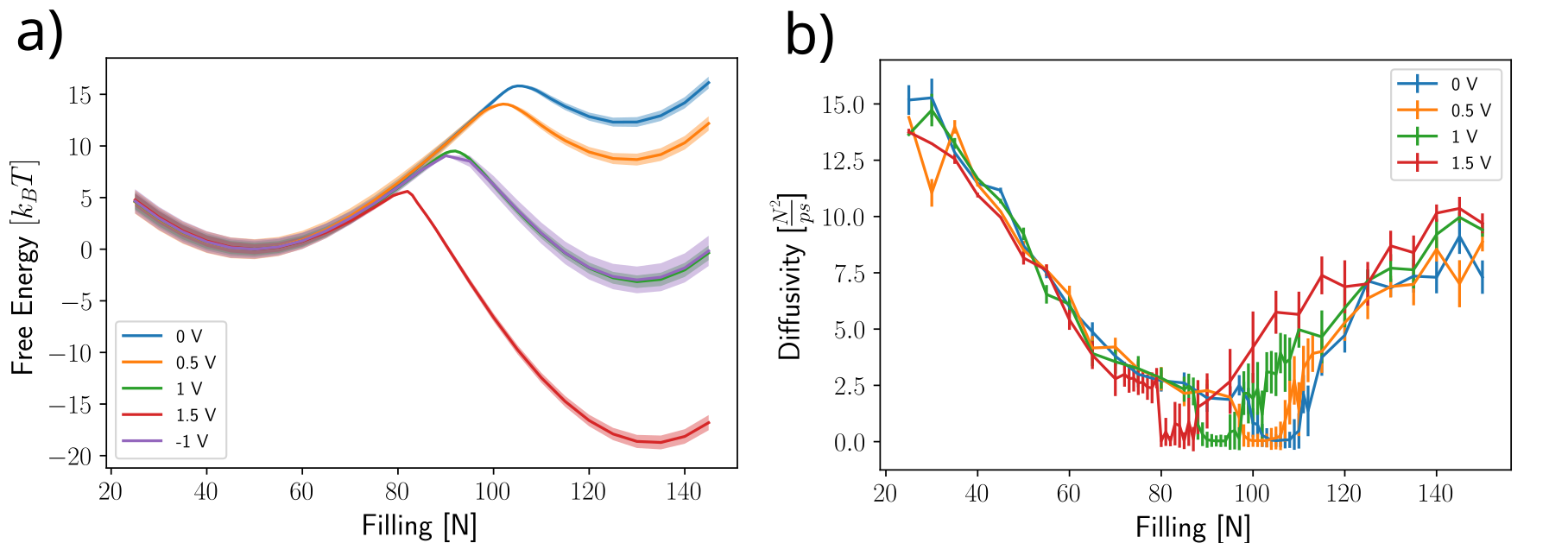}
    \caption{The effect of voltage on the free energy and the state dependent diffusivity.
    a) Free energy as a function of the water filling for different values of applied voltage. As higher voltage is applied, the wet state becomes more and more favourable. 
    Positive and negative applied voltages have the same effect, as can be seen from the overlap of the free energies at -1V and 1V b) Diffusivity as a function of the water filling for different voltages. For most values of the pore filling, more significantly close to the two minima (dry and wet states), the voltage does not significantly change the diffusivity. Differently, it can be noted how the minimum of the diffusivity shift to lower filling level at larger voltages, following the free energy barrier maximum. The shaded area on the free energy corresponds to the error associated with the integration computed starting from the free energy barrier, by using bootstrapping to estimate the error of the derivative of the free energy. Error bars on diffusivity represent the standard error associated with two different methods. 
    }
    \label{fig:free_RMD}
\end{figure*}

We computed the free energy $F(N,\Phi)$ and the state dependent diffusivity $D(N, \Phi)$ as a function of pore filling $N$ at different applied voltages $\Phi$ using RMD, 
see Fig.~\ref{fig:free_RMD}.
 
\begin{figure*}
    \centering
    \includegraphics[width=\linewidth]{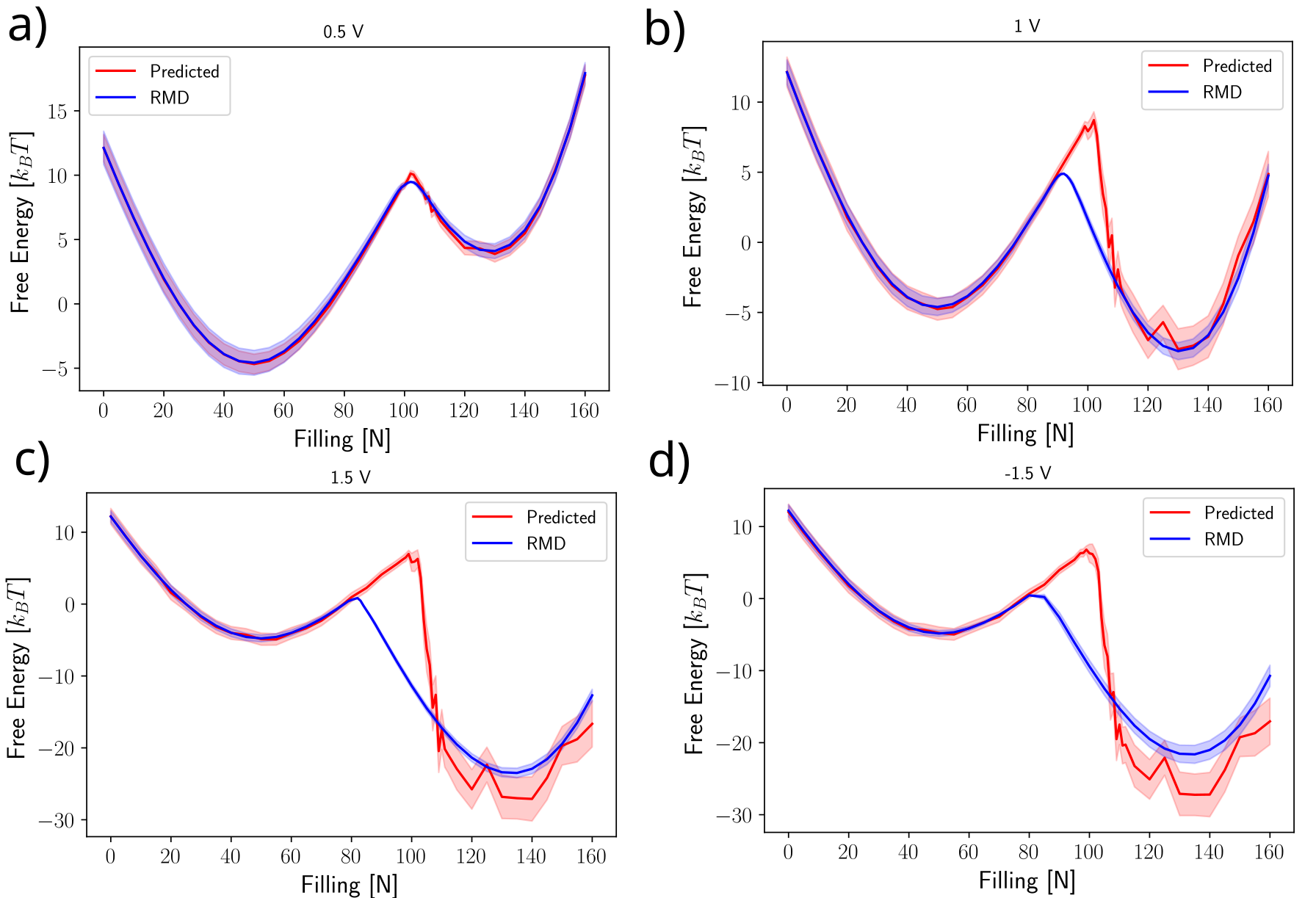}
    \caption{Comparison between free energy computed with RMD and prediction. 
    Panels a), b), c) and d) correspond to the free energies associated with $\Phi=$~0.5, 1, 1.5 and -1~V respectively. 
    For the lowest applied voltages, 0.5V, the predicted and computed values using RMD match, while for larger applied voltages both the barrier maximum and the free energy of the wet state are not well captured. The predicted lines (blue) use the values of P1 and P2 of Fig.~\ref{fig:displacementcharge}. 
    Shaded areas correspond to the propagation of the errors  95\% confidence interval computed using bootstrapping.
    }
    \label{fig:comparison_voltage_effect}
\end{figure*}

\begin{figure}[t]
    \centering
    \includegraphics[width=\linewidth]{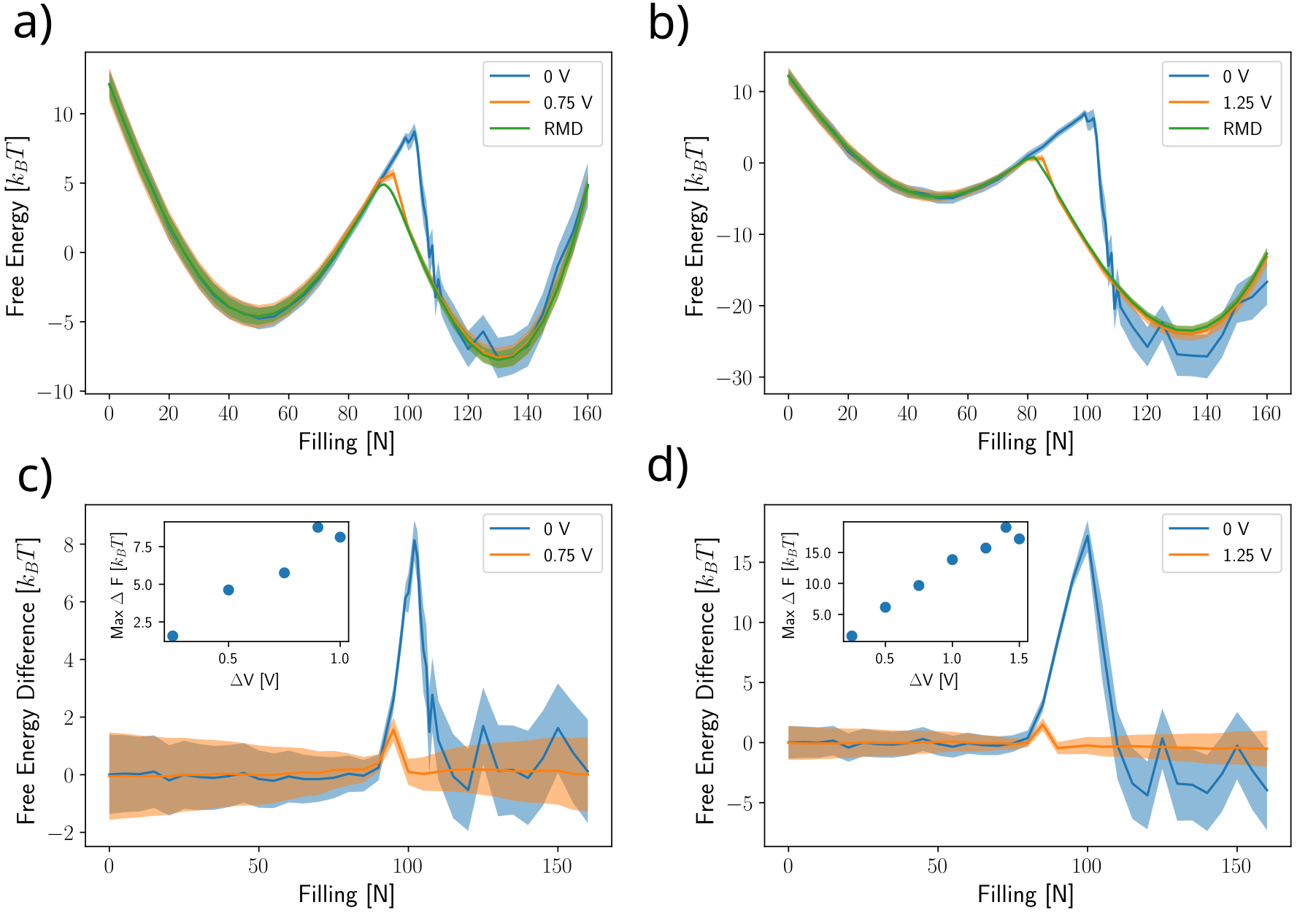}
    \caption{Predicting the free energy using a closer voltage point improves the accuracy of the model. Panel a) shows the predicting of the free energy profile at 1 V, using Eq.~\ref{eq:theory_free}, expanding around 0 V (blue) or around 0.75 V (orange). Panel b) shows the predicting of the free energy profile at 1.5 V, expanding around 0 V (blue) or expanding around 1.25 V (orange). Panel c) and shows the free energy difference between the free energy measured with RMD and the expansions at different voltages, for the case of 1 V, while panel d) corresponds to the case of 1.5 V. The insets in each panel show the maximum difference of the free energy considering expansions at different voltage differences.
    }
    \label{fig:comparison_voltage_effect_partial}
\end{figure}

We remark that the free energy of a dynamical system is an equilibrium (different from steady state) quantity, hence the protocol used here is strictly valid only in the absence of electric or mass currents.
The fluid in the system of Fig.~\ref{fig:system} does not include free ions, hence our system reaches equilibrium under every applied voltage and filling constraints. Thus, the quantity we are computing corresponds to the free energy of the system, despite the applied electric field.
Figure~\ref{fig:free_RMD}a shows that voltage has a non-linear effect on the free energy of the wet state. Applying $\Phi=0.5$~V changes the free energy difference between the wet and dry state by less than $5\,k_BT$ while $\Phi=1$~V changes this difference by more than $15\,k_BT$ and $\Phi=1.5$~V by more than $30\,k_BT$. 
The non-linear effect of voltage in wetting of hydrophobic pores has been shown previously~\cite{Trick2017} and it is in line with the prediction of our electrowetting theory, Eq.~\eqref{eq:theory_free} -- as it will be shown in Fig.~\ref{fig:comparison_voltage_effect} and~\ref{fig:comparison_voltage_effect_partial}. 
We note that the position of the free energy barrier changes with applied voltage, meaning that the volume (and possibly shape) of the critical nucleation bubble changes with applied voltage, requiring higher bubble volumes (corresponding to lower filling 
 $N$) to make the pore dry.
Panel b shows that the minimum of the diffusivity follows the maximum of the free energy barrier, shifting to lower filling levels as the voltage increases.
Instead, differently from what was previously shown for different pressures~\cite{Paulo2023JCP}, the magnitude of the diffusivity of the filling variable changes with applied voltage.

We then compared the free energy profiles computed using RMD and the ones predicted by the expansion presented in Eq.~\eqref{eq:theory_free} with respect to $\Phi_0=0$, see Fig.~\ref{fig:comparison_voltage_effect}.
Panels a-d show the theoretical prediction at different voltages, $\Delta \Phi= 0.5$, 1, 1.5 and -1 V, using $F_0(N,\Phi_0=0)$ and the values of $P_1(N,\Phi_0=0)$ and $P_2(N,\Phi_0=0)$ shown in Fig.~\ref{fig:displacementcharge}. 
A fair agreement is found for the lowest voltage, $\Delta \Phi = 0.5$~V. 
However, as expected, the discrepancy becomes larger  as the applied voltages increases. More specifically, the free energy differences between the wet and dry states are still well predicted at every voltage (within the estimated error) but the free energy barriers are significantly overestimated by Eq.~\eqref{eq:theory_free}. Moreover, the location of the free-energy maximum  does not match the correct filling level $N$ for $\Delta \Phi>0.5$~V.
Both simulations and the predicted energy profile show that positive and negative voltages have similar free energy profiles, see Fig.~\ref{fig:comparison_voltage_effect}b,d.

The main source of error in the prediction stems from the fact that both $P_1$ and $P_2$ depend on voltage. 
Indeed, it is possible to improve the accuracy of the free energy prediction  by computing $F$, $P_1$ and $P_2$ at voltages closer to the predicted one, which is equivalent to performing the Taylor expansion in Eq.~\eqref{eq:theory_free} at larger voltages, i.e., using $F_0=F_0(N,|\Phi_0|>0)$ and corresponding $P_1$ and $P_2$.
As an example, in Fig.~\ref{fig:comparison_voltage_effect_partial}a,b we compared the prediction of Eq.~\eqref{eq:theory_free} at $\Phi = 1$~V and 1.5~V, respectively, with that computed with RMD, starting from different $F_0$.
In panel a are shown the predictions of $F(N,\Phi=1V)$ by using $F_0$, $P_1$ and $P_2$ computed -- by RMD -- at $\Phi_0=0$~V (blue) and $\Phi_0=0.75$~V (orange);
in panel b the prediction for $F(N,\Phi=1.5V)$ is done by using $F_0$, $P_1$ and $P_2$ measured at $\Phi_0=0$~V (blue) and $\Phi_0=1.25$~V (orange).
In both cases, as expected, the theoretical expansion at smaller $\Delta \Phi = \Phi - \Phi_0$ gives much closer predictions to the the free energy computed by RMD.
Panels Fig.~\ref{fig:comparison_voltage_effect_partial}c,d quantify the differences between the profiles shown in the panels a and b, respectively, for each filling level $N$.
In both panels, it is possible to see that the largest error comes from the free energy barrier separating the wet and dry states. The insets in both panels show how the maximum difference between the free energies lowers as the theoretical expansion is performed closer to the predicted voltage, being as low as 2~$k_BT$ for a difference $\Delta \Phi \sim 0.25$~V.

\subsection{Effect of applied voltage on $P_1$ and $P_2$.}

\begin{figure}
    \centering
    \includegraphics[width=\linewidth]{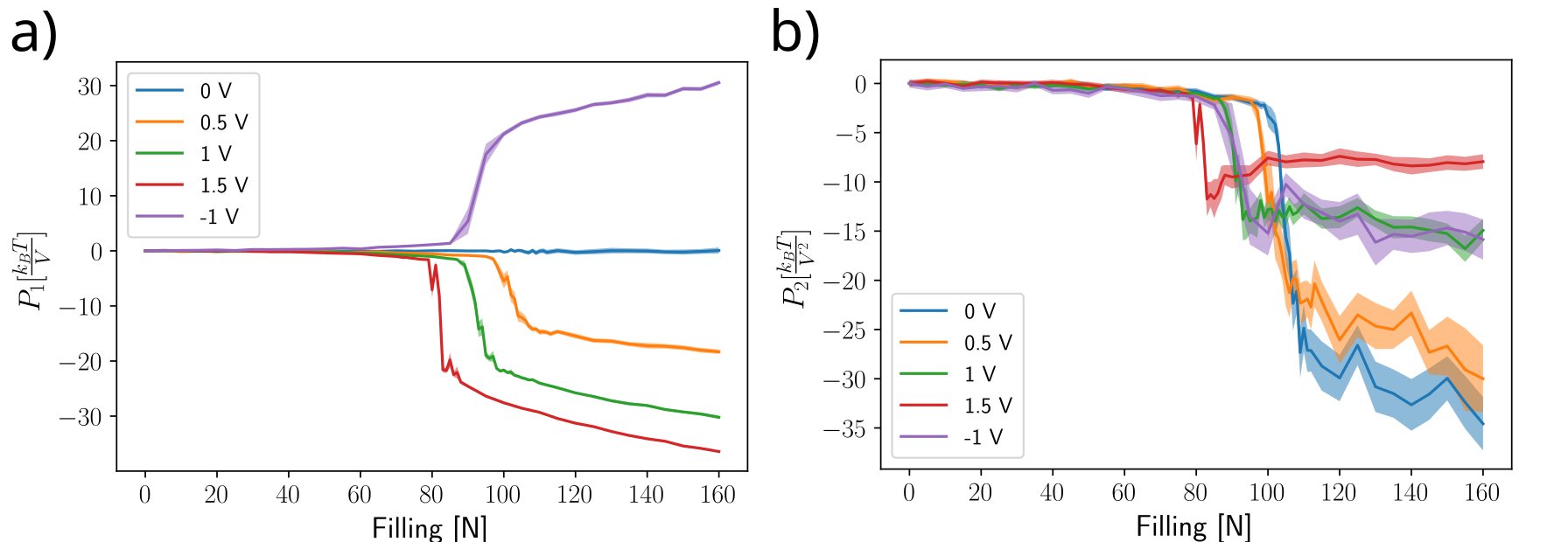}
    \caption{Dependence of $P_1$ and $P_2$ on the voltage. Panel a) shows the values of $P_1$ for different positive voltages and one negative voltage, highlighting its symmetric behaviour. Panel b) shows the value of $P_2$ for different positive voltages and one negative voltage, to highlight that positive and negative applied voltages give the same result. Both $P_1$ and $P_2$ have a step like behaviour, switching at the transition state.}
    \label{fig:p1_p2_voltage}
\end{figure}

Both $P_1$ and $P_2$ depend on the applied voltage, see Fig.~\ref{fig:p1_p2_voltage}. 
$P_1$ is symmetric with respect to the applied voltage, see the green ($1$~V) and violet ($-1$~V) lines in Fig.~\ref{fig:p1_p2_voltage}a.
The symmetry of $P_1$ comes from the symmetry of the system, hence positive and negative applied voltages must have the same effect on the free energy, i.e. it must be that:
$\Phi \; P_1(N,\Phi) = -\Phi \; P_1(N,-\Phi)$.
Moreover, $P_1(\Phi,N)$ cannot be everywhere constant for voltages different from 0. 
Indeed, consider the free energy profile of the system at $\Phi = 1$~V: in that case, applying a voltage difference of $+1$~V or $-1$~V must give different results, because in one case you would be considering the free energy profile at 2 V, while the other you would be considering the free energy profile at 0 V.

$P_1$ is related to the dipole moment of the liquid inside the pore, and the interpretation of Fig.~\ref{fig:p1_p2_voltage}a is that the polarization of the liquid phase is higher than that of the vapour phase, as previously discussed, with a discontinuous jump close to the transition state. It is also possible to observe an increase of $P_1$ in the wet state (as $N$ increases), with the slope depending on the applied voltage. Indeed, a larger electric field will make the water molecules more ordered, and so the total dipole moment of the liquid inside the pore will increase with the voltage; moreover, the number of ordered dipoles increases with the number of water molecules filling the pore. 

On the other hand, $P_2$ has the same value for positive and negative voltages, see Fig.~\ref{fig:p1_p2_voltage}b. 
As previously discussed, $P_2$ has the units of a capacitance and is observed to increase with the filling $N$ in the wet state and to decrease with the applied voltage. Using the same analogy of the capacitor, this must mean that the dielectric constant of the medium goes down (as the distance and area of the plates remains the same). This result is in line with previously reported data from atomistic simulations in hydrophobic nanopores, showing that the relative permittivity of water decreases with applied electric field~\cite{Dzubiella2005}; this is linked to the disruption of the hydrogen bond network under confinement.

As we previously demonstrated, for this particular system, our second order expansion leads to errors lower than ca. $5\,{k_BT}$ on the barrier only for $\Delta \Phi \leq 0.5$~V; this error is not negligible when predicting the wetting and drying rates of an hydrophobically gated switch. 
However, it is impractical to compute $P_1$ and $P_2$ for multiple applied voltages, because this requires computationally expensive RMD simulations. 
Because $P_1 = \frac{dF}{d\Phi}$, we tried to find a phenomenological formula for $P_1(N,\Phi)$, which would allow us to compute directly the free energy at any $\Phi$ by integrating the expression.

\begin{figure}
    \centering
    \includegraphics[width=\linewidth]{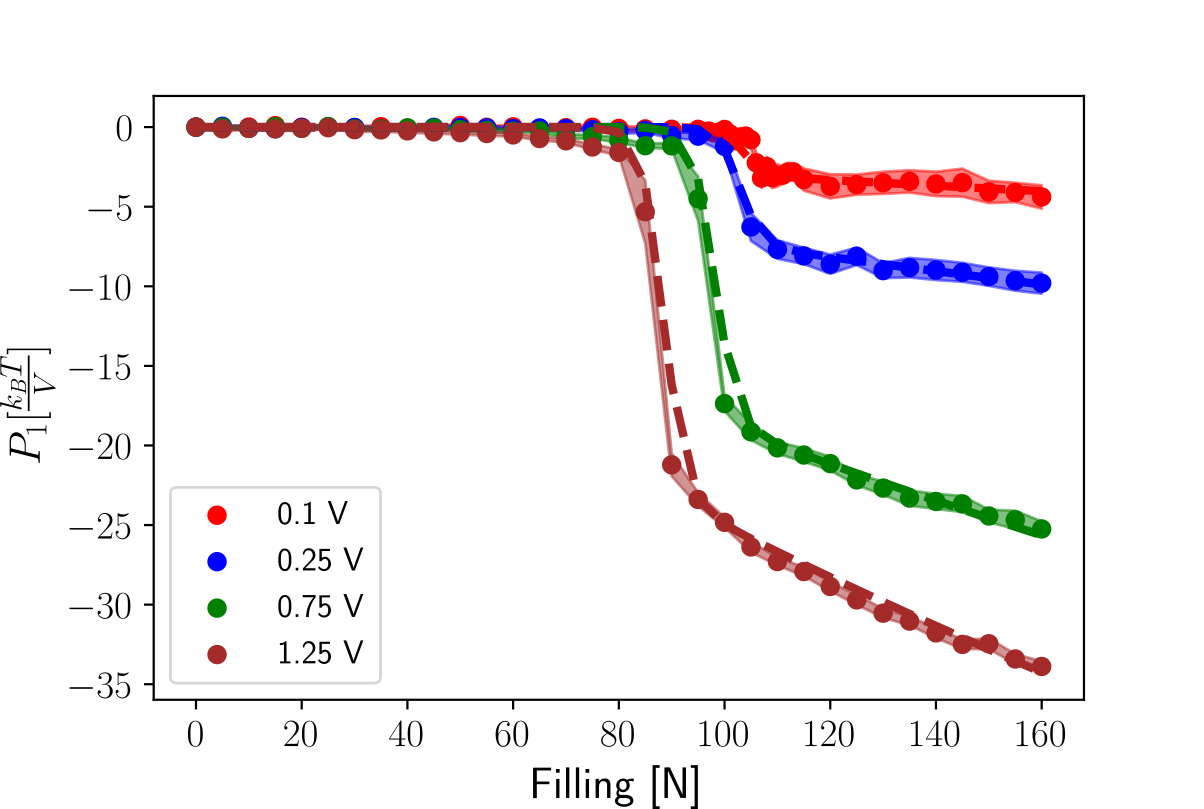}
    \caption{Accuracy of phenomenological function for $P_1$ and. The phenomenological function describing $P_1$ (dashed line) gives good agreement with the values computed with RMD (dots), for voltages not used for the fit.}
    \label{fig:phenomenology}
\end{figure}

To model $P_1$ we consider that its expression should contain some switching function that changes from a low value to a high value at a critical filling level $N^*$. This critical value depends on the applied voltage, and so we write it as $N^*(\Phi)$. The switching function chosen by us is a sigmoid, with expression 
$\frac{a(\Phi)}{1+e^{-1/2(N-N^*(\Phi))}}$, with $a(\Phi)$ representing a voltage-dependent amplitude. This expression does not completely capture the behaviour of $P_1$, in particular, its increase in absolute value at higher fillings, see Fig.~\ref{fig:p1_p2_voltage}a. In order to capture this trend, one can multiply the described switching function by $N^{2/3}$. This left us to determine the expression of $a(\Phi)$, $b(\Phi)$, and $N^*(\Phi)$, 
which we modelled as $a = a_0 \tanh(\Phi)$, $b = 1/2$, and $N^*=N_0 - c|\Phi|$. 
The final expression for $P_1$ has the form:
\begin{equation}
    P_1(N,\Phi) = \frac{a_0 \tanh(\Phi)N^{2/3}}{1+e^{-\frac{(N-N_0+c|\Phi|)}{2}}}, 
\end{equation}
see the supplementary materials for the fitting involved in the determination of the appropriate constants, $N_0$, $c$, $a_0$. The agreement between that expression and voltages not used in the fitting is shown in Fig.~\ref{fig:phenomenology}.

The fitted phenomenological expression is rigorously only applicable to the specific system shown in Fig.~\ref{fig:system}.
Currently, this expression is not dimensionally coherent, and we believe this expression is not a general one, especially for more complex systems, but it reasonably describes the shapes of $P_1$ at different voltages. 
Further work is needed to clarify if a more general, physically informed expression of this function exists. 
We believe that the arguments presented here can anyway help as guidance to fit similar phenomenological expressions in other systems, effectively speeding up the estimation of the free energy at any voltage using a minimal set of RMD simulations.

\begin{figure}
    \centering
    \includegraphics[width=\linewidth]{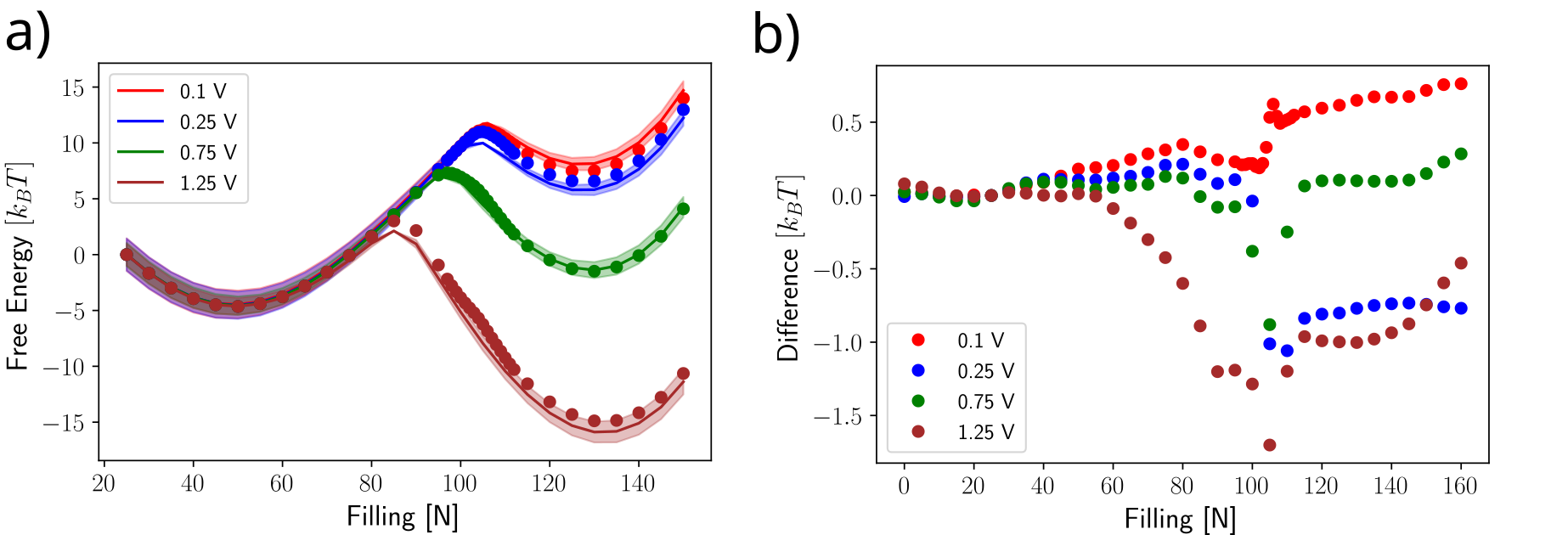}
    \caption{Free energy profile obtained from integrating $P_1$. The free energy profiles, panel a), computed using these expressions (full lines) are very accurate when compared with that computed using RMD (dots). The difference between the predicted and the simulated free energy, panel b), is always less than 2 $k_BT$, with the largest deviation located closed to the free energy barrier.}
    \label{fig:phenomenology_free}
\end{figure}

Having access to an analytical form of $P_1$ means that we can reconstruct the free energy profile at any voltage, by integrating $P_1$ and adding this value to the free energy at $\Phi=0$, see Fig.~\ref{fig:phenomenology_free}a. This approach leads to a much more accurate estimation of the free energy profile than that of Fig.~\ref{fig:comparison_voltage_effect}, with the maximum error at free energy barrier being less than 2~$k_BT$, see \tr{Fig.~\ref{fig:phenomenology_free}b}. Despite the fact that this protocol can be used to generate the free energy profile at arbitrary voltages, we believe that extrapolations higher than $\Delta \Phi \gg 1.5$~V will most likely give unphysical results.

\subsection{Estimating the wetting and drying rates}

\begin{figure}
    \centering
    \includegraphics[width=\linewidth]{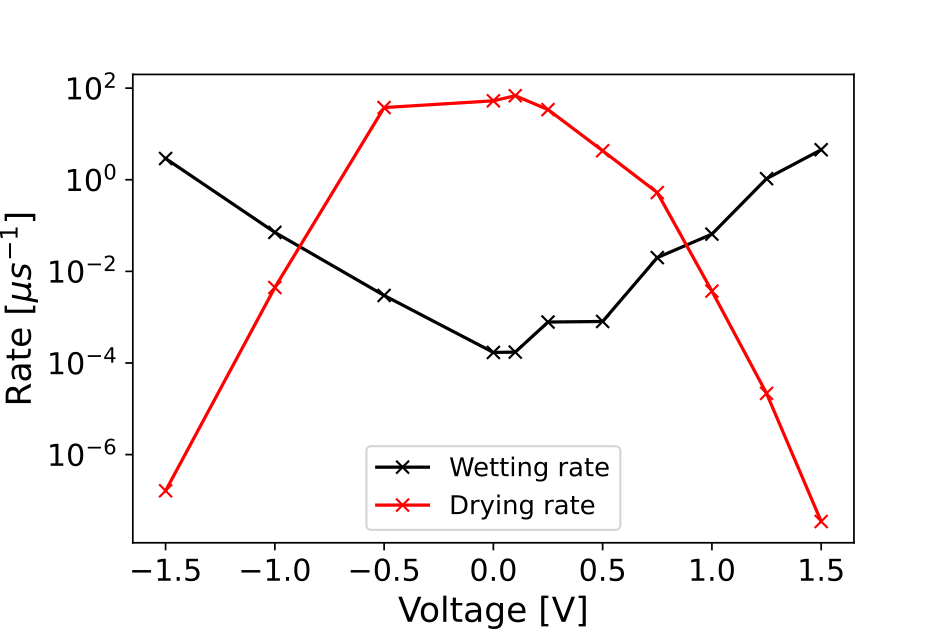}
    \caption{Wetting and drying rates at different voltages. Both the drying and the wetting rate change by multiple orders of magnitude. When no voltage is applied, the drying rate is 6 orders of magnitude faster than the wetting rate, while when 1.5 V is applied to the system, the wetting rate is 6 orders of magnitude faster than the drying rate. }
    \label{fig:rates}
\end{figure}

With the idea of iontronic switches in mind, we compute the wetting and drying times at different voltages by using a rate theory expression~\cite{Hnggi1990} fed with the free energy profile $F(N,\Phi)$ and the state dependent diffusivity $D(N,\Phi)$:
\begin{equation}
    t_{w}(\Phi) = \int_{N_d}^{N_w}\frac{e^{\frac{F(N,\Phi)}{k_BT}}}{D(N,\Phi)}dN \;\int_{N_d}^{N_w}e^{\frac{-F(N,\Phi)}{k_BT}}\left(\int_{N_d}^{N}\frac{e^{\frac{F(N,\Phi)}{k_BT}}}{D(N,\Phi)}dN'\right) dN,
    \label{eq:ratetheory}
\end{equation}
with $t_{w}$ being the wetting time and $N_d$ and $N_w$ the filling levels of the dry and wet states. To compute the drying time, one just needs to switch the integration limits. The wetting and drying rates are then computed as the inverse of the wetting and drying times, respectively.
We plot the results of these calculations in Fig. \ref{fig:rates}. We observe a complete inversion of the gating behaviour when voltage is applied. At low applied voltages, the system dries, i.e., switches off, much faster than it spontaneously wets, meaning that it mostly stays in the dry state. As voltage increases the wetting rate increases, but it is still lower than $1~\mu s^{-1}$ up to an applied voltage of 1~V. For higher voltages, the drying rate becomes much smaller than the wetting rate, because the probability of the system being in the wet state is given by $P_w=\frac{r_w}{r_w+r_d}$, where $r_w$ and $r_d$ are the wetting and dryng rates. The wet state of the pore is conductive to ions, while the dry state is not, which means that voltage can be used to switch the system from a conductive to a non-conductive state, and vice-versa.

\section{Conclusions}

In this theoretical and computational work, we  present a framework to predict the wetting and drying behaviour of hydrophobic nanopores under an applied electric potential. Hydrophobic nanopores can be used as simple switches in nanofluidic devices because they promote the formation of a bubble that can stop the ion flux. The present framework could be used to study the behaviour of other hydrophobically gated nanopores, like ion channels and biological pores.

We show that it is possible to build a theory of the effect of voltage on pore wetting, which provides the free energy at arbitrary applied voltages. By informing such theory by molecular dynamics simulations we were able to compute the wetting free energy profiles at different voltages.
Simulations at zero voltage are used to compute the coefficients $P_1$ and $P_2$ that enter into the mathematical expansion of the free energy, described by Eq.~\ref{eq:theory_free}. As expected, this expansion works best for values of applied voltage smaller than 0.5 V, but still predicts correctly the probability of finding the system in either the wet or dry state; the estimate of the free energy barriers, instead, becomes inaccurate as the voltage departs from zero.

Because the coefficients $P_1$ and $P_2$ depend on voltage, we computed them at different applied voltages and fit a phenomenological expression for $P_1$, which we could then use to accurately estimate the free energy at any intermediate value of applied voltage. The predictions are accurate even for voltages that were not used in the fitting of the phenomenological expression, with an error less than 2~$k_BT$ at every filling point.

We also computed the state dependent diffusivity for the filling variable $N$ at different values of applied voltage. In this way, we were able to estimate the wetting and drying rates at different applied voltages. At low applied voltages, the pore dries in the GHz range, while it wets only in the Hz range. We observed an inversion at high voltages, where the drying rate is 6 orders of magnitude lower than the wetting rate, while at no applied voltage the drying rate is 6 orders of magnitude larger. This means that at high voltages, the pores wet in the GHz range and dry in the Hz range. 

The proposed framework is general and can be used to predict the voltage-controlled switching behaviour of more complex hydrophobic nanopores of interest in iontronic applications. Recently, this protocol has been used to predict the memristive behaviour associated with hydrophobic gating in both computational and experimental settings~\cite{paulomem2023}, showing promise as a powerful tool for the design of novel nanofluidic devices.

\section{Acknowledgement}

This research is part of a project that has received funding from the European Research Council (ERC) under the European Union's Horizon 2020 research and innovation programme (grant agreement No. 803213). 
We acknowledge the EuroHPC Joint Undertaking for awarding this project access to the EuroHPC supercomputer LUMI, hosted by CSC (Finland) and the LUMI consortium through a EuroHPC Regular Access call.

\section*{Data Availability Statement}

The data that support the findings of this study are openly available in Zenodo at 10.5281/zenodo.8412939

\bibliography{aipsamp}

\end{document}